\pdfoutput=1

\documentclass[aps,prapplied,twocolumn,superscriptaddress]{revtex4-1}

\usepackage{url}
\usepackage{amsmath}
\usepackage{graphicx}
\usepackage{color}
\usepackage{float}
\usepackage{soul}

\usepackage{hyperref}
\usepackage{braket}
\usepackage{siunitx}
\usepackage{bm}

\usepackage[T1]{fontenc}

\renewcommand{\d}[1]{\ensuremath{\operatorname{d}\!{#1}}}
\newcommand{\subscript}[1]{$_{\text{#1}}$}
\newcommand{\superscript}[1]{$^{\text{#1}}$}
\newcommand{\yso}[0]{Y\subscript{2}SiO\subscript{5}}
\newcommand{\nd}[1]{\superscript{\!{#1}}Nd}

\newcommand{\comment}[1]{} 

\begin{document}


\title{High-cooperativity coupling of a rare-earth spin ensemble to a superconducting resonator using yttrium orthosilicate as a substrate}


\author{Gavin Dold}
\affiliation{London Centre for Nanotechnology, University College London, London WC1H 0AH, United Kingdom}
\affiliation{National Physical Laboratory, Hampton Road, Teddington TW11 0LW, United Kingdom}

\author{Christoph W. Zollitsch}
\affiliation{London Centre for Nanotechnology, University College London, London WC1H 0AH, United Kingdom}

\author{James O'Sullivan}
\affiliation{London Centre for Nanotechnology, University College London, London WC1H 0AH, United Kingdom}

\author{Sacha Welinski}
\affiliation{Universit\'e PSL, Chimie ParisTech, CNRS, Institut de Recherche de Chimie Paris, 75005 Paris, France}

\author{Alban Ferrier}
\affiliation{Universit\'e PSL, Chimie ParisTech, CNRS, Institut de Recherche de Chimie Paris, 75005 Paris, France}
\affiliation{Facult\'e des Sciences et Ing\'enierie,  Sorbonne Universit\'e, 75005 Paris, France}

\author{Philippe Goldner}
\affiliation{Universit\'e PSL, Chimie ParisTech, CNRS, Institut de Recherche de Chimie Paris, 75005 Paris, France}

\author{S.E. de Graaf}
\affiliation{National Physical Laboratory, Hampton Road, Teddington TW11 0LW, United Kingdom}

\author{Tobias Lindstr\"om}
\affiliation{National Physical Laboratory, Hampton Road, Teddington TW11 0LW, United Kingdom}

\author{John J. L. Morton}
\affiliation{London Centre for Nanotechnology, University College London, London WC1H 0AH, United Kingdom}
\affiliation{Department of Electronic and Electrical Engineering, UCL, London WC1E 7JE, United Kingdom}


\date{\today}

\begin{abstract}
Yttrium orthosilicate (\yso{}, or YSO) has proved to be a convenient host for rare-earth ions used in demonstrations of microwave quantum memories and optical memories with microwave interfaces, and shows promise for coherent microwave--optical conversion owing to its favourable optical and spin properties. The strong coupling required by such microwave applications could be achieved using superconducting resonators patterned directly on \yso{}, and hence we investigate here the use of \yso{} as an alternative to sapphire or silicon substrates for superconducting hybrid device fabrication. A NbN resonator with frequency 6.008 GHz and low power quality factor $Q \approx \num{400000}$ was fabricated on a \yso{} substrate doped with isotopically enriched \nd{145}. Measurements of dielectric loss yield a loss-tangent $\tan\delta = \num{4e-6}$, comparable to sapphire. Electron spin resonance (ESR) measurements performed using the resonator show the characteristic angular dependence expected from the anisotropic \nd{145} spin, and the coupling strength between resonator and electron spins is in the high cooperativity regime ($C = 30$). These results demonstrate \yso{} as an excellent substrate for low-loss, high-Q microwave resonators, especially in applications for coupling to optically-accessible rare earth spins.
\end{abstract}

\maketitle

\section{Introduction}

Rare-earth ions (REIs) in crystals are promising candidate systems for quantum information applications as they possess electron and nuclear spins as well as optical transitions at telecom wavelengths. Coherence times in these systems range from milliseconds \cite{Wolfowicz2015} for microwave excitations stored in an electron spin to hours \cite{Zhong2015,Rancic2017} for optically accessible nuclear excitations. REIs typically possess large g-factors (beneficial for strong coupling), a variety of nuclear spin states, and offer opportunities for optical pumping. Such properties have led to proposals for using rare-earth ions as a multimode microwave quantum memory \cite{Probst2015}, a photonic memory for a quantum repeater \cite{Sangouard2011}, and as a microwave--optical transducer \cite{Williamson2014,Fernandez-Gonzalvo2015} for use in quantum networks \cite{Kimble2008}.  


Yttrium orthosilicate (\yso{}) is a widely used crystalline host for REIs in such quantum information applications, as its constituent elements provide an environment with a low background of nuclear magnetic moments that would otherwise contribute to spin decoherence and inhomogeneous broadening. The narrow homogeneous linewidths, for example down to \SI{3.5}{kHz} for the \SI{883}{nm} transition \cite{Usmani2010} in Nd\superscript{3+}:\yso{}, have been exploited as an optical quantum memory \cite{Hedges2010,Jobez2015} enabling storage of entangled states \cite{Clausen2011} and teleportation with \SI{93}{\percent} fidelity \cite{Bussieres2014}.

\begin{figure}
	\includegraphics[width = 0.99\columnwidth]{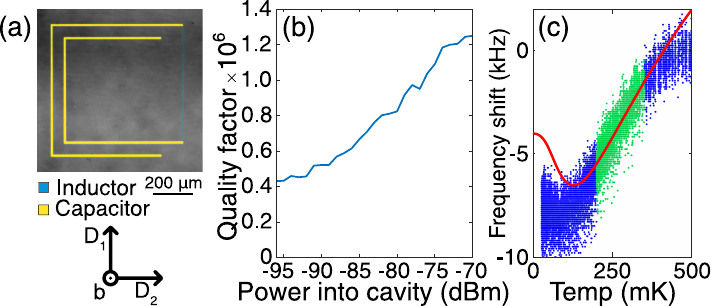}%
	\caption{\label{fig:fig1}(a) False-colour micrograph of the superconducting thin-ring resonator on \nd{145}:\yso{} substrate. (b) Power dependence of the loaded quality factor at zero field. At high powers the two-level systems (TLS) begin to saturate, resulting in an increase in $Q$ \cite{Zmuidzinas2012}. (c) Frequency shift of resonator as function of temperature, from which a fit to theory \cite{Lindstrom2009} (red) of the gradient of the linear section (green) yields the loss-tangent $\tan\delta = \num{4e-6}$.}
\end{figure}

While optical storage experiments using REIs in \yso{} can make use of a suitably optically dense medium, for microwave storage a cavity is employed to achieve a coupling between an ensemble of REIs and the microwave cavity field, so that excitations can be coherently exchanged between the two. 
This can be achieved with 3D microwave cavities~\cite{Fernandez-Gonzalvo2015}, which offer homogeneous $B_1$ fields and spatial mode-matching between microwave and optical fields, or planar superconducting resonators with small mode volumes and quality factors over $10^5$ to yield high spin-number sensitivities~\cite{Bienfait2015} and act as an interface between superconducting qubits and spin ensembles~\cite{Kubo2011}.
Dielectric resonators fabricated from rare-earth doped crystals such as YAG (Y$_3$Al$_5$O$_{12}$) \cite{Farr2015} have used resonant modes with $Q \approx \num{e4}$ to drive ESR transitions with cooperativity $C \approx 600$, while planar superconducting resonators have been coupled to rare-earths ($Q \approx \num{e3}$, $C=36$, \cite{Probst2013}), group V donors in silicon~\cite{Bienfait2015,Zollitsch2015}, NV centres in diamond~\cite{Kubo2011}, and ruby (Al\subscript{2}O\subscript{3}:Cr\superscript{3+}) \cite{Schuster2010}.

\begin{figure*}[!t]
	\includegraphics[width = 0.99\textwidth]{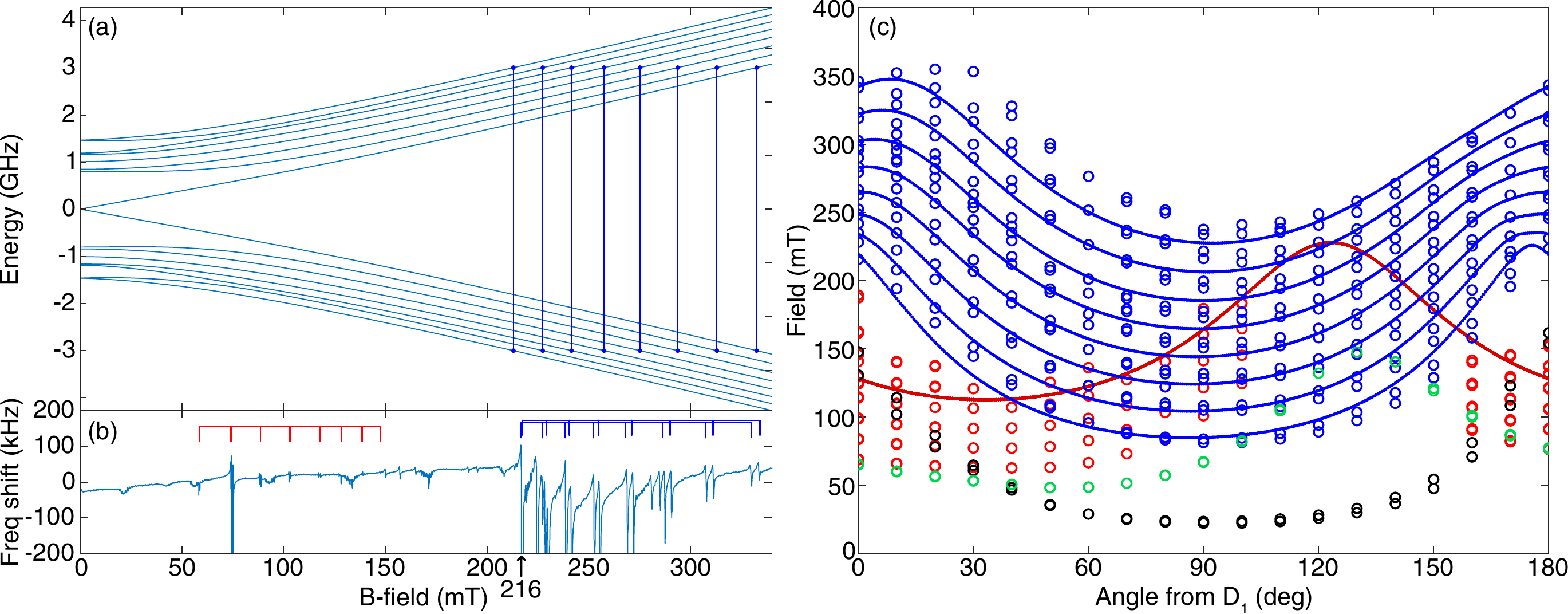}
	\caption{\label{fig:fig2}(a) Breit-Rabi diagram of \nd{145} in \yso{} for magnetic field B\subscript{0} oriented along D\subscript{1}, with ESR transitions at \SI{6.008}{GHz} indicated. (b) ESR spectrum with the magnetic field B\subscript{0} oriented along D\subscript{1}, as measured with a VNA tracking the resonator centre frequency. The quadratic dependence of resonator frequency with applied magnetic field has been subtracted. Transitions from Site~1 are indicated in blue, Site~2 in red. Two sets of site 1 transitions are seen due to a broken sub-site degeneracy. The avoided crossing studied in Section III.B is indicated at $B_0 = \SI{216}{mT}$. (c) Roadmap of ESR transition fields with respect to the B\subscript{0} angle from the D\subscript{1} axis, in the D\subscript{1}--D\subscript{1} plane. Circles mark the positions of transitions from traces such as in (b), with Site~1 in blue and Site~2 in red. Signals from site 2 between \ang{110}--\ang{150} could not be resolved due to weak transition amplitudes at these angles. Lines represent simulated data assuming an ideal rotation about the crystal b axis. For Site~2 we only show one line as only the g-tensor is known for this site. Green and black represent as yet unidentified impurities.}
\end{figure*}

Experiments coupling superconducting resonators to doped crystals have typically used a \emph{flip--chip} approach with the sample glued or mechanically pressed onto the resonator chip \cite{Probst2013}. This has the drawback of complicating the fabrication process, creates an additional interface layer between the resonator and sample increasing dielectric losses from spurious two-level systems (TLS) in the interface \cite{Burnett2014}, as well as a variable gap between device and spins which is less controllable than fabrication on the substrate itself.

Common substrates for superconducting resonators include silicon and sapphire (Al\subscript{2}O\subscript{3}) \cite{Oliver2013}. Dielectric losses in these are usually quantified by the loss tangent $\tan\delta$, describing the ratio between the imaginary and real components of the complex permittivity of the material. This creates a limit on the quality factor achievable in superconducting resonators $Q = 1 /\tan\delta$. For high-Q devices, sapphire is often preferred for its low loss-tangent \cite{Lindstrom2009} $\tan\delta < \num{e-5}$, leading to quality factors approaching $10^6$.

In this work we investigate the suitability of using \yso{} itself as a substrate for fabrication of planar superconducting devices. We fabricate a resonator on the polished surface of a Nd-doped \yso{} sample, and find the device has a Q-factor $>10^5$ at low powers due to its low loss-tangent $\tan\delta = \num{4e-6}$. We observe an anisotropic ESR spectrum matching that from simulations, and measure the coupling strength between resonator and electron spin to be in the high-cooperativity regime. These results suggest fabrication on doped \yso{} is compatible with high-Q devices while simultaneously enabling a coupling to crystals doped with rare-earth ions, and could be a promising route toward scalable hybrid superconductor--spin quantum circuits.

\section{Device}

The device is a lumped-element superconducting resonator, fabricated on a Czochralski-grown single crystal of \yso{} \cite{Shoudu1999} doped with \SI{10}{ppm} isotopically purified \superscript{145}Nd. The crystal was cut into a \SI{5}{mm}$\times$\SI{5}{mm}$\times$\SI{460}{\micro\metre} chip along the principal dielectric axes (D\subscript{1}, D\subscript{2}, b) and a face perpendicular to b was polished for thin-film growth. \yso{} has two inequivalent crystal sites where a Y\superscript{$3+$} ion can be substituted by a rare-earth RE\superscript{$3+$} ion. The large ionic radius of Nd\superscript{3+} results in the larger crystal Site~1 being preferentially populated\cite{Wolfowicz2015}, resulting in a stronger signal from Site~1 over Site~2. Due to the crystal's C$^{6}_{2\textrm{h}}$ (C$_2$/c) space group, each site has two orientations related by a $\pi$ rotation around the crystal b axis. These two orientations are termed \emph{sub-sites}, and their ESR properties are degenerate for $B_0$ fields applied in the D\subscript{1}--D\subscript{2} plane or parallel to the b axis.

 The resonator consists of a \SI{2}{\micro\metre} wire which functions as an inductor due to the constriction increasing the contribution of kinetic inductance, and a pair of \SI{10}{\micro\metre} thick capacitive arms separated by a \SI{50}{\micro\metre} gap. The use of narrow features increases the resilience of the resonator to applied magnetic field, and the large gap between the capacitive arms reduces the peak electric field of the resonator decreasing its susceptibility to dielectric losses from two-level systems. This forms a lumped-element resonator seen in the micrograph in Fig.~\ref{fig:fig1} which generates an oscillating $B_1$ magnetic field around the inductor which can drive ESR transitions. Details of this resonator design's properties in applied magnetic field will be described in a future publication.

Fabrication consisted of \SI{40}{nm} of sputtered NbN being patterned by photolithography and a SF\subscript{6}/Ar reactive ion etch process. The patterned chip was enclosed within a 3D copper cavity ($\mathrm{Q} \approx 100$) to suppress spontaneous emission from the resonator to the environment \cite{Bienfait2015}. This was installed in the bore of a vector magnet in a dilution refrigerator at \SI{10}{mK} (measured at the mixing chamber plate) and probed using a vector network analyser (VNA) at zero applied magnetic field.

The lumped-element resonator has a frequency of \SI{6.008}{GHz} and an asymmetric lineshape due to interference with the background transmission of the 3D cavity. Fitting with a Fano resonance \cite{Barnthaler2010} yields a quality factor $Q \approx \num{400000}$ in the low-power limit in Fig.~\ref{fig:fig1}. By tracking the centre frequency versus temperature \cite{Zmuidzinas2012} we find the dielectric losses in the device are comparable to resonators fabricated on sapphire \cite{Pappas2011,Lindstrom2009} with a loss-tangent $\tan\delta = \num{4e-6}$, where the filling factor $F \approx \tfrac{1}{2}$ representing the fraction of magnetic field penetrating the substrate has been factored out.

This demonstrates that \yso{} is well-suited for devices incorporating resonators and superconducting qubits which are typically susceptible to dielectric losses from TLSs \cite{Faoro2006}, while also incorporating doped spins for cavity QED.

\section{Electron spin resonance}

\subsection{ESR spectrum}

Having studied the resonator, we now use it to perform ESR to characterise the Nd spins in the \yso{} that are coupled to the resonator.
A magnetic field $B_0$ was applied in the plane of the superconducting thin film, roughly perpendicular to the \yso{} b crystal axis, to minimise the magnetic flux threaded through the superconductor. The field was swept up to 360 mT and the resonator was used to observe ESR transitions, monitoring changes to its centre frequency tracked with a VNA. By this method a series of spectra were taken as a function of $B_0$ orientation in the D\subscript{1}--D\subscript{2} plane to extract the angular dependence, or ``roadmap'', of ESR line positions, in order to confirm the presence of Nd spins in the substrate and measure their coupling to the resonator.

For an electron spin $\bm{S}$ coupled to a nuclear spin $\bm{I}$ the Hamiltonian accounting for the electron Zeeman and hyperfine terms is $H =  \mu_\mathrm{B} \bm{B}^\mathrm{T}\hspace*{-3pt}\bm{g} \hat{\bm{S}} + \hat{\bm{S}}^\mathrm{T}\hspace*{-3pt}\bm{A} \hat{\bm{I}}$. The anisotropy of \yso{} results in the parameters $\bm{g}$ and $\bm{A}$ being tensors, previously calculated from spectroscopic studies of site 1 of $^{145}$\nd{}:\yso{} at \SI{9.4}{GHz} \cite{Maier-Flaig2013} as
\begin{align*}
\bm{g} &= \begin{pmatrix}
1.30& 0.62& 0.22\\
0.62& -2.07& 1.62\\
0.22& 1.62& -2.86
\end{pmatrix}_{\mathrm{(D_1,D_2,b)}}
\\
\bm{A} &= \begin{pmatrix}
-37.1& -99.9& -83.4\\
-99.9& -589.2& 169.4\\
-83.4 & 169.4& -678.4
\end{pmatrix}_{\mathrm{(D_1,D_2,b)}} \si{\mega\hertz}
\end{align*}

A characteristic ESR spectrum ($B_0$ along the D\subscript{1} axis) is plotted in Fig.~\ref{fig:fig2}(b). We observe a series of eight ESR transitions indicated in Fig.~\ref{fig:fig2}(a), corresponding to the \superscript{145}Nd nuclear spin \big($I=\tfrac{7}{2}$\big) in Site~1. A small misalignment between the D\subscript{1}--D\subscript{2} and superconductor planes breaks the degeneracy between the crystal sub-sites and causes a further splitting of each line into two.

The roadmap of ESR spectrum with respect to $B_0$-field angle is plotted in Fig.~\ref{fig:fig2}(c), with transitions marked by circles. The angular dependence seen in these spectra matches well simulations for Nd spins in Site~1 (blue) while we also observe a signal at lower magnetic fields (higher g-factor) which we attribute to Site~2 (red). The full spin Hamiltonian of \superscript{145}Nd in Site~2 including hyperfine terms has not been reported, therefore our simulations only account for Zeeman term (i.e.~g tensor). Additional unidentified impurities are evident from the different angular dependences of 2--3 transitions marked in green and black. 

These measurements enable ESR transitions to be positively identified by comparing their angular dependence to simulations \cite{Wisby2016} and bulk ESR data, ensuring further analysis of spin--resonator coupling properties can be clearly linked to a particular spin species.

\subsection{High-cooperativity coupling}
\begin{figure}[!t]
	\includegraphics[width = 0.99\columnwidth]{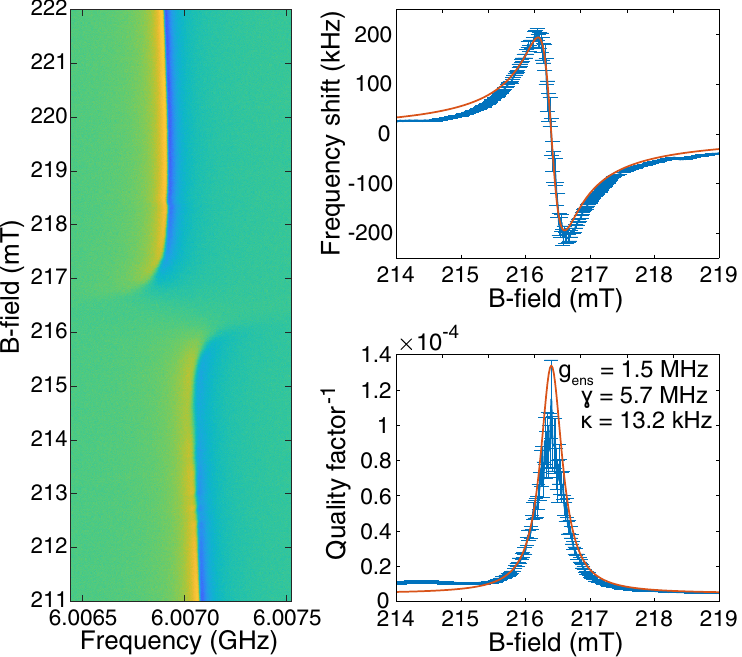}
	\caption{\label{fig:fig3}Left: onset of an avoided crossing when sweeping B-field through resonance. Right: resonator frequency shift and quality factor as a function of B-field strength when sweeping through resonance. The uncertainty in resonator quality factor increases at the centre of the transition due to a reduction in prominence of the resonance. A simultaneous fit to all resonances as a function of magnetic field, using the equations in the main text, yields a dependence of frequency and quality factor with field shown in red, and a high-cooperativity coupling $C=\num{30}$.}
\end{figure}

The highest intensity ESR transition for \superscript{145}Nd in Site~1 with $m_\mathrm{I}=\tfrac{7}{2}$ at $B_0 = \SI{216}{mT}$ along D\subscript{1} was selected for an evaluation of the spin--resonator coupling strength. Measuring the resonator S\subscript{21} with a VNA while sweeping magnetic field strength $B$, we observe the onset of an avoided crossing shown in Fig.~\ref{fig:fig3}. Fitting the resonator frequency $\omega$ and linewidth $\kappa$ at each field point\cite{Abe2011} allows us to extract the coupling strength $g_\textrm{ens}$, inhomogeneous spin ensemble half-width $\gamma_\mathrm{s}$, and resonator half-width $\kappa_\mathrm{c}$ with
\begin{align*}
\omega = \omega_\mathrm{c} - g_\mathrm{ens}^2 \Delta / (\Delta^2 + \gamma_\mathrm{s}^2)\\
\kappa = \kappa_\mathrm{c} - g_\mathrm{ens}^2 \gamma_\mathrm{s} / (\Delta^2 + \gamma_\mathrm{s}^2)
\end{align*}
where $\Delta = m_0 (B - B_\mathrm{0})/\hbar$ is the field detuning calculated from the spin magnetic moment $m_0 = \hbar \frac{\d{\omega}}{\d{B_0}}$. This accounts for cases where $\frac{\d{\omega}}{\d{B}}$ varies with $B$, as is the case for mixed spin systems in the low-field limit or near zero first-order Zeeman (ZEFOZ) points.

From this we extract $g_\mathrm{ens} = \SI{1.5}{MHz}$, $\gamma_\mathrm{s} = \SI{5.7}{MHz}$, and $\kappa = \SI{13.2}{kHz}$. This corresponds to a high cooperativity $C = \tfrac{g_\mathrm{ens}^2}{\kappa \gamma} = 30$.

For a bulk doped sample there is no clear single-spin coupling rate as the $\tfrac{1}{r}$ decay of the field strength from the wire results in a $B_1$ inhomogeneity spanning orders of magnitude while all spins in the sample contribute to the signal. However, a characteristic rate representing the average coupling strength over the spins contributing \SI{50}{\percent} of the measured signal can be calculated. This corresponds to a region within \SI{4}{\micro\metre} of the wire with $N\approx\num{6e8}$ resonant \superscript{145}Nd spins in the same sub-site, accounting for the thermal population in the $m_\mathrm{I}=\tfrac{7}{2}$ ESR transition at \SI{10}{mK}. From this we derive a characteristic single-spin coupling rate $g_0 = \tfrac{g_\textrm{ens}}{\sqrt{N}}\approx\SI{50}{Hz}$ leading to an expected Purcell enhanced emission rate\cite{Bienfait2015a} of $\Gamma_\textrm{P} = \tfrac{4g_0^2}{\kappa} \approx \SI{1}{Hz}$.


These results indicate the coupling regime is limited by the spin linewidth $\gamma_\mathrm{s} > g_\textrm{ens}$, or equivalently that the Nd spin ensemble is decohering at a faster rate than excitations are being exchanged between it and the resonator. Steps toward improving this coupling strength could include using a more strongly doped rare-earth sample to increase $g_\textrm{ens}$ though this could also increase $\gamma_\mathrm{s}$, or exploiting the coherence-enhancing ZEFOZ transitions in REIs \cite{Ortu2018} to decrease $\gamma_\mathrm{s}$. The significant $B_1$ inhomogeneity in this resonator design also poses challenges for performing coherent operations on the entire spin ensemble with pulsed ESR. This could be mitigated by coupling to implanted layers of REIs \cite{Kukharchyk2014,Wisby2016}, or by using resonators designed to generate a more homogeneous $B_1$ field.

\section{Conclusion}

We investigated the use of \yso{} as an alternative to sapphire or silicon substrates for the fabrication of superconducting devices. The fabricated NbN lumped-element resonator had a quality factor $Q \approx \num{400000}$ and a loss-tangent $\tan\delta = \num{4e-6}$ comparable to sapphire \cite{Lindstrom2009}, and yielded a high-cooperativity coupling $C = 30$ between the resonator and a $^{145}$\nd{} spin ensemble. These results demonstrate \yso{} is well-suited for superconducting devices, while also enabling an interaction with optically-accessible rare earth spins 

Further studies of this device include performing pulsed ESR measurements in the high-cooperativity regime and measuring coherence properties of the spin ensemble. The methods shown here are also applicable to other REIs in \yso{}, for example Er with its \SI{1540}{nm} telecom band optical transition \cite{Fernandez-Gonzalvo2015} and Yb which exhibits a large oscillator strength \cite{Welinski2016} and coherence-enhancing ZEFOZ and near-ZEFOZ transitions \cite{Ortu2018}. It may also prove interesting to study the suitability as a substrate of other crystalline hosts for REIs, such as yttrium aluminium garnet (YAG), yttrium lithium fluoride (YLF), yttrium orthovanadate (YVO$_4$), and calcium tungstate (CaWO$_4$).

Fabricating superconducting devices on crystals doped with REIs as demonstrated here shows promise for integrating these optical elements alongside the fast information processing available from superconducting qubits. This is important step toward making a microwave--optical transducer capable of connecting such quantum processors within a quantum network. By exploiting long coherence times available from REIs this is also a route to integrating fast quantum electronics with a spin ensemble acting as a quantum memory.

\begin{acknowledgments}
This work has received funding from the Engineering and Physical Sciences Research Council (EPSRC) through the Centre for Doctoral Training in Delivering Quantum Technologies (Grant No.~EP/L015242/1), the European Community's Seventh Framework Programme Grant No.~279781 (ASCENT), the IMTO Cancer AVIESAN (Cancer Plan, C16027HS, MALT), and the UK government's Department for Business, Energy and Industrial Strategy.
\end{acknowledgments}

\bibliography{bib}

\begin{thebibliography}{34}%
\makeatletter
\providecommand \@ifxundefined [1]{%
 \@ifx{#1\undefined}
}%
\providecommand \@ifnum [1]{%
 \ifnum #1\expandafter \@firstoftwo
 \else \expandafter \@secondoftwo
 \fi
}%
\providecommand \@ifx [1]{%
 \ifx #1\expandafter \@firstoftwo
 \else \expandafter \@secondoftwo
 \fi
}%
\providecommand \natexlab [1]{#1}%
\providecommand \enquote  [1]{``#1''}%
\providecommand \bibnamefont  [1]{#1}%
\providecommand \bibfnamefont [1]{#1}%
\providecommand \citenamefont [1]{#1}%
\providecommand \href@noop [0]{\@secondoftwo}%
\providecommand \href [0]{\begingroup \@sanitize@url \@href}%
\providecommand \@href[1]{\@@startlink{#1}\@@href}%
\providecommand \@@href[1]{\endgroup#1\@@endlink}%
\providecommand \@sanitize@url [0]{\catcode `\\12\catcode `\$12\catcode
  `\&12\catcode `\#12\catcode `\^12\catcode `\_12\catcode `\%12\relax}%
\providecommand \@@startlink[1]{}%
\providecommand \@@endlink[0]{}%
\providecommand \url  [0]{\begingroup\@sanitize@url \@url }%
\providecommand \@url [1]{\endgroup\@href {#1}{\urlprefix }}%
\providecommand \urlprefix  [0]{URL }%
\providecommand \Eprint [0]{\href }%
\providecommand \doibase [0]{https://doi.org/}%
\providecommand \selectlanguage [0]{\@gobble}%
\providecommand \bibinfo  [0]{\@secondoftwo}%
\providecommand \bibfield  [0]{\@secondoftwo}%
\providecommand \translation [1]{[#1]}%
\providecommand \BibitemOpen [0]{}%
\providecommand \bibitemStop [0]{}%
\providecommand \bibitemNoStop [0]{.\EOS\space}%
\providecommand \EOS [0]{\spacefactor3000\relax}%
\providecommand \BibitemShut  [1]{\csname bibitem#1\endcsname}%
\let\auto@bib@innerbib\@empty
\bibitem [{\citenamefont {Wolfowicz}\ \emph {et~al.}(2015)\citenamefont
  {Wolfowicz}, \citenamefont {Maier-Flaig}, \citenamefont {Marino},
  \citenamefont {Ferrier}, \citenamefont {Vezin}, \citenamefont {Morton},\ and\
  \citenamefont {Goldner}}]{Wolfowicz2015}%
  \BibitemOpen
  \bibfield  {author} {\bibinfo {author} {\bibfnamefont {G.}~\bibnamefont
  {Wolfowicz}}, \bibinfo {author} {\bibfnamefont {H.}~\bibnamefont
  {Maier-Flaig}}, \bibinfo {author} {\bibfnamefont {R.}~\bibnamefont {Marino}},
  \bibinfo {author} {\bibfnamefont {A.}~\bibnamefont {Ferrier}}, \bibinfo
  {author} {\bibfnamefont {H.}~\bibnamefont {Vezin}}, \bibinfo {author}
  {\bibfnamefont {J.~J.~L.}\ \bibnamefont {Morton}},\ and\ \bibinfo {author}
  {\bibfnamefont {P.}~\bibnamefont {Goldner}},\ }\bibfield  {title} {\bibinfo
  {title} {{Coherent Storage of Microwave Excitations in Rare-Earth Nuclear
  Spins}},\ }\bibfield  {journal} {\bibinfo  {journal} {Physical Review
  Letters}\ }\href {https://doi.org/10.1103/PhysRevLett.114.170503}
  {10.1103/PhysRevLett.114.170503} (\bibinfo {year} {2015})\BibitemShut
  {NoStop}%
\bibitem [{\citenamefont {Zhong}\ \emph {et~al.}(2015)\citenamefont {Zhong},
  \citenamefont {Hedges}, \citenamefont {Ahlefeldt}, \citenamefont
  {Bartholomew}, \citenamefont {Beavan}, \citenamefont {Wittig}, \citenamefont
  {Longdell},\ and\ \citenamefont {Sellars}}]{Zhong2015}%
  \BibitemOpen
  \bibfield  {author} {\bibinfo {author} {\bibfnamefont {M.}~\bibnamefont
  {Zhong}}, \bibinfo {author} {\bibfnamefont {M.~P.}\ \bibnamefont {Hedges}},
  \bibinfo {author} {\bibfnamefont {R.~L.}\ \bibnamefont {Ahlefeldt}}, \bibinfo
  {author} {\bibfnamefont {J.~G.}\ \bibnamefont {Bartholomew}}, \bibinfo
  {author} {\bibfnamefont {S.~E.}\ \bibnamefont {Beavan}}, \bibinfo {author}
  {\bibfnamefont {S.~M.}\ \bibnamefont {Wittig}}, \bibinfo {author}
  {\bibfnamefont {J.~J.}\ \bibnamefont {Longdell}},\ and\ \bibinfo {author}
  {\bibfnamefont {M.~J.}\ \bibnamefont {Sellars}},\ }\bibfield  {title}
  {\bibinfo {title} {{Optically addressable nuclear spins in a solid with a
  six-hour coherence time}},\ }\href {https://doi.org/10.1038/nature14025}
  {\bibfield  {journal} {\bibinfo  {journal} {Nature}\ }\textbf {\bibinfo
  {volume} {517}},\ \bibinfo {pages} {177} (\bibinfo {year}
  {2015})}\BibitemShut {NoStop}%
\bibitem [{\citenamefont {Ran{\v{c}}i{\'{c}}}\ \emph
  {et~al.}(2017)\citenamefont {Ran{\v{c}}i{\'{c}}}, \citenamefont {Hedges},
  \citenamefont {Ahlefeldt},\ and\ \citenamefont {Sellars}}]{Rancic2017}%
  \BibitemOpen
  \bibfield  {author} {\bibinfo {author} {\bibfnamefont {M.}~\bibnamefont
  {Ran{\v{c}}i{\'{c}}}}, \bibinfo {author} {\bibfnamefont {M.~P.}\ \bibnamefont
  {Hedges}}, \bibinfo {author} {\bibfnamefont {R.~L.}\ \bibnamefont
  {Ahlefeldt}},\ and\ \bibinfo {author} {\bibfnamefont {M.~J.}\ \bibnamefont
  {Sellars}},\ }\bibfield  {title} {\bibinfo {title} {{Coherence time of over a
  second in a telecom-compatible quantum memory storage material}},\ }\href
  {https://doi.org/10.1038/nphys4254} {\bibfield  {journal} {\bibinfo
  {journal} {Nature Physics}\ }\textbf {\bibinfo {volume} {14}},\ \bibinfo
  {pages} {50} (\bibinfo {year} {2017})}\BibitemShut {NoStop}%
\bibitem [{\citenamefont {Probst}\ \emph {et~al.}(2015)\citenamefont {Probst},
  \citenamefont {Rotzinger}, \citenamefont {Ustinov},\ and\ \citenamefont
  {Bushev}}]{Probst2015}%
  \BibitemOpen
  \bibfield  {author} {\bibinfo {author} {\bibfnamefont {S.}~\bibnamefont
  {Probst}}, \bibinfo {author} {\bibfnamefont {H.}~\bibnamefont {Rotzinger}},
  \bibinfo {author} {\bibfnamefont {A.~V.}\ \bibnamefont {Ustinov}},\ and\
  \bibinfo {author} {\bibfnamefont {P.~A.}\ \bibnamefont {Bushev}},\ }\bibfield
   {title} {\bibinfo {title} {{Microwave multimode memory with an erbium spin
  ensemble}},\ }\href {https://doi.org/10.1103/PhysRevB.92.014421} {\bibfield
  {journal} {\bibinfo  {journal} {Physical Review B}\ }\textbf {\bibinfo
  {volume} {92}},\ \bibinfo {pages} {014421} (\bibinfo {year}
  {2015})}\BibitemShut {NoStop}%
\bibitem [{\citenamefont {Sangouard}\ \emph {et~al.}(2011)\citenamefont
  {Sangouard}, \citenamefont {Simon}, \citenamefont {de~Riedmatten},\ and\
  \citenamefont {Gisin}}]{Sangouard2011}%
  \BibitemOpen
  \bibfield  {author} {\bibinfo {author} {\bibfnamefont {N.}~\bibnamefont
  {Sangouard}}, \bibinfo {author} {\bibfnamefont {C.}~\bibnamefont {Simon}},
  \bibinfo {author} {\bibfnamefont {H.}~\bibnamefont {de~Riedmatten}},\ and\
  \bibinfo {author} {\bibfnamefont {N.}~\bibnamefont {Gisin}},\ }\bibfield
  {title} {\bibinfo {title} {{Quantum repeaters based on atomic ensembles and
  linear optics}},\ }\href {https://doi.org/10.1103/RevModPhys.83.33}
  {\bibfield  {journal} {\bibinfo  {journal} {Reviews of Modern Physics}\
  }\textbf {\bibinfo {volume} {83}},\ \bibinfo {pages} {33} (\bibinfo {year}
  {2011})}\BibitemShut {NoStop}%
\bibitem [{\citenamefont {Williamson}\ \emph {et~al.}(2014)\citenamefont
  {Williamson}, \citenamefont {Chen},\ and\ \citenamefont
  {Longdell}}]{Williamson2014}%
  \BibitemOpen
  \bibfield  {author} {\bibinfo {author} {\bibfnamefont {L.~A.}\ \bibnamefont
  {Williamson}}, \bibinfo {author} {\bibfnamefont {Y.~H.}\ \bibnamefont
  {Chen}},\ and\ \bibinfo {author} {\bibfnamefont {J.~J.}\ \bibnamefont
  {Longdell}},\ }\bibfield  {title} {\bibinfo {title} {{Magneto-optic modulator
  with unit quantum efficiency}},\ }\href
  {https://doi.org/10.1103/PhysRevLett.113.203601} {\bibfield  {journal}
  {\bibinfo  {journal} {Physical Review Letters}\ }\textbf {\bibinfo {volume}
  {113}},\ \bibinfo {pages} {1} (\bibinfo {year} {2014})}\BibitemShut {NoStop}%
\bibitem [{\citenamefont {Fernandez-Gonzalvo}\ \emph
  {et~al.}(2015)\citenamefont {Fernandez-Gonzalvo}, \citenamefont {Chen},
  \citenamefont {Yin}, \citenamefont {Rogge},\ and\ \citenamefont
  {Longdell}}]{Fernandez-Gonzalvo2015}%
  \BibitemOpen
  \bibfield  {author} {\bibinfo {author} {\bibfnamefont {X.}~\bibnamefont
  {Fernandez-Gonzalvo}}, \bibinfo {author} {\bibfnamefont {Y.~H.}\ \bibnamefont
  {Chen}}, \bibinfo {author} {\bibfnamefont {C.}~\bibnamefont {Yin}}, \bibinfo
  {author} {\bibfnamefont {S.}~\bibnamefont {Rogge}},\ and\ \bibinfo {author}
  {\bibfnamefont {J.~J.}\ \bibnamefont {Longdell}},\ }\bibfield  {title}
  {\bibinfo {title} {{Coherent frequency up-conversion of microwaves to the
  optical telecommunications band in an Er:YSO crystal}},\ }\href
  {https://doi.org/10.1103/PhysRevA.92.062313} {\bibfield  {journal} {\bibinfo
  {journal} {Physical Review A - Atomic, Molecular, and Optical Physics}\
  }\textbf {\bibinfo {volume} {92}},\ \bibinfo {pages} {1} (\bibinfo {year}
  {2015})}\BibitemShut {NoStop}%
\bibitem [{\citenamefont {Kimble}(2008)}]{Kimble2008}%
  \BibitemOpen
  \bibfield  {author} {\bibinfo {author} {\bibfnamefont {H.~J.}\ \bibnamefont
  {Kimble}},\ }\bibfield  {title} {\bibinfo {title} {{The quantum internet}},\
  }\href {https://doi.org/10.1038/nature07127} {\bibfield  {journal} {\bibinfo
  {journal} {Nature}\ }\textbf {\bibinfo {volume} {453}},\ \bibinfo {pages}
  {1023} (\bibinfo {year} {2008})},\ \Eprint {https://arxiv.org/abs/0806.4195}
  {arXiv:0806.4195} \BibitemShut {NoStop}%
\bibitem [{\citenamefont {Usmani}\ \emph {et~al.}(2010)\citenamefont {Usmani},
  \citenamefont {Afzelius}, \citenamefont {de~Riedmatten},\ and\ \citenamefont
  {Gisin}}]{Usmani2010}%
  \BibitemOpen
  \bibfield  {author} {\bibinfo {author} {\bibfnamefont {I.}~\bibnamefont
  {Usmani}}, \bibinfo {author} {\bibfnamefont {M.}~\bibnamefont {Afzelius}},
  \bibinfo {author} {\bibfnamefont {H.}~\bibnamefont {de~Riedmatten}},\ and\
  \bibinfo {author} {\bibfnamefont {N.}~\bibnamefont {Gisin}},\ }\bibfield
  {title} {\bibinfo {title} {{Mapping multiple photonic qubits into and out of
  one solid-state atomic ensemble}},\ }\href
  {https://doi.org/10.1038/ncomms1010} {\bibfield  {journal} {\bibinfo
  {journal} {Nature Communications}\ }\textbf {\bibinfo {volume} {1}},\
  \bibinfo {pages} {1} (\bibinfo {year} {2010})}\BibitemShut {NoStop}%
\bibitem [{\citenamefont {Hedges}\ \emph {et~al.}(2010)\citenamefont {Hedges},
  \citenamefont {Longdell}, \citenamefont {Li},\ and\ \citenamefont
  {Sellars}}]{Hedges2010}%
  \BibitemOpen
  \bibfield  {author} {\bibinfo {author} {\bibfnamefont {M.~P.}\ \bibnamefont
  {Hedges}}, \bibinfo {author} {\bibfnamefont {J.~J.}\ \bibnamefont
  {Longdell}}, \bibinfo {author} {\bibfnamefont {Y.}~\bibnamefont {Li}},\ and\
  \bibinfo {author} {\bibfnamefont {M.~J.}\ \bibnamefont {Sellars}},\
  }\bibfield  {title} {\bibinfo {title} {{Efficient quantum memory for
  light}},\ }\href {https://doi.org/10.1038/nature09081} {\bibfield  {journal}
  {\bibinfo  {journal} {Nature}\ }\textbf {\bibinfo {volume} {465}},\ \bibinfo
  {pages} {1052} (\bibinfo {year} {2010})}\BibitemShut {NoStop}%
\bibitem [{\citenamefont {Jobez}\ \emph {et~al.}(2015)\citenamefont {Jobez},
  \citenamefont {Laplane}, \citenamefont {Timoney}, \citenamefont {Gisin},
  \citenamefont {Ferrier}, \citenamefont {Goldner},\ and\ \citenamefont
  {Afzelius}}]{Jobez2015}%
  \BibitemOpen
  \bibfield  {author} {\bibinfo {author} {\bibfnamefont {P.}~\bibnamefont
  {Jobez}}, \bibinfo {author} {\bibfnamefont {C.}~\bibnamefont {Laplane}},
  \bibinfo {author} {\bibfnamefont {N.}~\bibnamefont {Timoney}}, \bibinfo
  {author} {\bibfnamefont {N.}~\bibnamefont {Gisin}}, \bibinfo {author}
  {\bibfnamefont {A.}~\bibnamefont {Ferrier}}, \bibinfo {author} {\bibfnamefont
  {P.}~\bibnamefont {Goldner}},\ and\ \bibinfo {author} {\bibfnamefont
  {M.}~\bibnamefont {Afzelius}},\ }\bibfield  {title} {\bibinfo {title}
  {{Coherent Spin Control at the Quantum Level in an Ensemble-Based Optical
  Memory}},\ }\href {https://doi.org/10.1103/PhysRevLett.114.230502} {\bibfield
   {journal} {\bibinfo  {journal} {Physical Review Letters}\ }\textbf {\bibinfo
  {volume} {114}},\ \bibinfo {pages} {230502} (\bibinfo {year}
  {2015})}\BibitemShut {NoStop}%
\bibitem [{\citenamefont {Clausen}\ \emph {et~al.}(2011)\citenamefont
  {Clausen}, \citenamefont {Usmani}, \citenamefont {Bussi{\`{e}}res},
  \citenamefont {Sangouard}, \citenamefont {Afzelius}, \citenamefont
  {de~Riedmatten},\ and\ \citenamefont {Gisin}}]{Clausen2011}%
  \BibitemOpen
  \bibfield  {author} {\bibinfo {author} {\bibfnamefont {C.}~\bibnamefont
  {Clausen}}, \bibinfo {author} {\bibfnamefont {I.}~\bibnamefont {Usmani}},
  \bibinfo {author} {\bibfnamefont {F.}~\bibnamefont {Bussi{\`{e}}res}},
  \bibinfo {author} {\bibfnamefont {N.}~\bibnamefont {Sangouard}}, \bibinfo
  {author} {\bibfnamefont {M.}~\bibnamefont {Afzelius}}, \bibinfo {author}
  {\bibfnamefont {H.}~\bibnamefont {de~Riedmatten}},\ and\ \bibinfo {author}
  {\bibfnamefont {N.}~\bibnamefont {Gisin}},\ }\bibfield  {title} {\bibinfo
  {title} {{Quantum storage of photonic entanglement in a crystal.}},\ }\href
  {https://doi.org/10.1038/nature09662} {\bibfield  {journal} {\bibinfo
  {journal} {Nature}\ }\textbf {\bibinfo {volume} {469}},\ \bibinfo {pages}
  {508} (\bibinfo {year} {2011})}\BibitemShut {NoStop}%
\bibitem [{\citenamefont {Bussieres}\ \emph {et~al.}(2014)\citenamefont
  {Bussieres}, \citenamefont {Clausen}, \citenamefont {Tiranov}, \citenamefont
  {Korzh}, \citenamefont {Verma}, \citenamefont {Nam}, \citenamefont {Marsili},
  \citenamefont {Ferrier}, \citenamefont {Goldner}, \citenamefont {Herrmann},
  \citenamefont {Silberhorn}, \citenamefont {Sohler}, \citenamefont
  {Afzelius},\ and\ \citenamefont {Gisin}}]{Bussieres2014}%
  \BibitemOpen
  \bibfield  {author} {\bibinfo {author} {\bibfnamefont {F.}~\bibnamefont
  {Bussieres}}, \bibinfo {author} {\bibfnamefont {C.}~\bibnamefont {Clausen}},
  \bibinfo {author} {\bibfnamefont {A.}~\bibnamefont {Tiranov}}, \bibinfo
  {author} {\bibfnamefont {B.}~\bibnamefont {Korzh}}, \bibinfo {author}
  {\bibfnamefont {V.~B.}\ \bibnamefont {Verma}}, \bibinfo {author}
  {\bibfnamefont {S.~W.}\ \bibnamefont {Nam}}, \bibinfo {author} {\bibfnamefont
  {F.}~\bibnamefont {Marsili}}, \bibinfo {author} {\bibfnamefont
  {A.}~\bibnamefont {Ferrier}}, \bibinfo {author} {\bibfnamefont
  {P.}~\bibnamefont {Goldner}}, \bibinfo {author} {\bibfnamefont
  {H.}~\bibnamefont {Herrmann}}, \bibinfo {author} {\bibfnamefont
  {C.}~\bibnamefont {Silberhorn}}, \bibinfo {author} {\bibfnamefont
  {W.}~\bibnamefont {Sohler}}, \bibinfo {author} {\bibfnamefont
  {M.}~\bibnamefont {Afzelius}},\ and\ \bibinfo {author} {\bibfnamefont
  {N.}~\bibnamefont {Gisin}},\ }\bibfield  {title} {\bibinfo {title} {{Quantum
  teleportation from a telecom-wavelength photon to a solid-state quantum
  memory}},\ }\href {https://doi.org/10.1038/nphoton.2014.215} {\bibfield
  {journal} {\bibinfo  {journal} {Nature Photonics}\ }\textbf {\bibinfo
  {volume} {8}},\ \bibinfo {pages} {775} (\bibinfo {year} {2014})}\BibitemShut
  {NoStop}%
\bibitem [{\citenamefont {Zmuidzinas}(2012)}]{Zmuidzinas2012}%
  \BibitemOpen
  \bibfield  {author} {\bibinfo {author} {\bibfnamefont {J.}~\bibnamefont
  {Zmuidzinas}},\ }\bibfield  {title} {\bibinfo {title} {{Superconducting
  Microresonators: Physics and Applications}},\ }\href
  {https://doi.org/10.1146/annurev-conmatphys-020911-125022} {\bibfield
  {journal} {\bibinfo  {journal} {Annual Review of Condensed Matter Physics}\
  }\textbf {\bibinfo {volume} {3}},\ \bibinfo {pages} {169} (\bibinfo {year}
  {2012})}\BibitemShut {NoStop}%
\bibitem [{\citenamefont {Lindstr{\"{o}}m}\ \emph {et~al.}(2009)\citenamefont
  {Lindstr{\"{o}}m}, \citenamefont {Healey}, \citenamefont {Colclough},
  \citenamefont {Muirhead},\ and\ \citenamefont {Tzalenchuk}}]{Lindstrom2009}%
  \BibitemOpen
  \bibfield  {author} {\bibinfo {author} {\bibfnamefont {T.}~\bibnamefont
  {Lindstr{\"{o}}m}}, \bibinfo {author} {\bibfnamefont {J.~E.}\ \bibnamefont
  {Healey}}, \bibinfo {author} {\bibfnamefont {M.~S.}\ \bibnamefont
  {Colclough}}, \bibinfo {author} {\bibfnamefont {C.~M.}\ \bibnamefont
  {Muirhead}},\ and\ \bibinfo {author} {\bibfnamefont {A.~Y.}\ \bibnamefont
  {Tzalenchuk}},\ }\bibfield  {title} {\bibinfo {title} {{Properties of
  superconducting planar resonators at millikelvin temperatures}},\ }\href
  {https://doi.org/10.1103/PhysRevB.80.132501} {\bibfield  {journal} {\bibinfo
  {journal} {Physical Review B - Condensed Matter and Materials Physics}\
  }\textbf {\bibinfo {volume} {80}},\ \bibinfo {pages} {2} (\bibinfo {year}
  {2009})}\BibitemShut {NoStop}%
\bibitem [{\citenamefont {Bienfait}\ \emph {et~al.}(2015)\citenamefont
  {Bienfait}, \citenamefont {Pla}, \citenamefont {Kubo}, \citenamefont {Stern},
  \citenamefont {Zhou}, \citenamefont {Lo}, \citenamefont {Weis}, \citenamefont
  {Schenkel}, \citenamefont {Thewalt}, \citenamefont {Vion}, \citenamefont
  {Esteve}, \citenamefont {Julsgaard}, \citenamefont {Moelmer}, \citenamefont
  {Morton},\ and\ \citenamefont {Bertet}}]{Bienfait2015}%
  \BibitemOpen
  \bibfield  {author} {\bibinfo {author} {\bibfnamefont {A.}~\bibnamefont
  {Bienfait}}, \bibinfo {author} {\bibfnamefont {J.~J.}\ \bibnamefont {Pla}},
  \bibinfo {author} {\bibfnamefont {Y.}~\bibnamefont {Kubo}}, \bibinfo {author}
  {\bibfnamefont {M.}~\bibnamefont {Stern}}, \bibinfo {author} {\bibfnamefont
  {X.}~\bibnamefont {Zhou}}, \bibinfo {author} {\bibfnamefont {C.~C.}\
  \bibnamefont {Lo}}, \bibinfo {author} {\bibfnamefont {C.~D.}\ \bibnamefont
  {Weis}}, \bibinfo {author} {\bibfnamefont {T.}~\bibnamefont {Schenkel}},
  \bibinfo {author} {\bibfnamefont {M.~L.~W.}\ \bibnamefont {Thewalt}},
  \bibinfo {author} {\bibfnamefont {D.}~\bibnamefont {Vion}}, \bibinfo {author}
  {\bibfnamefont {D.}~\bibnamefont {Esteve}}, \bibinfo {author} {\bibfnamefont
  {B.}~\bibnamefont {Julsgaard}}, \bibinfo {author} {\bibfnamefont
  {K.}~\bibnamefont {Moelmer}}, \bibinfo {author} {\bibfnamefont {J.~J.~L.}\
  \bibnamefont {Morton}},\ and\ \bibinfo {author} {\bibfnamefont
  {P.}~\bibnamefont {Bertet}},\ }\bibfield  {title} {\bibinfo {title}
  {{Reaching the quantum limit of sensitivity in electron spin resonance}},\
  }\href {https://doi.org/10.1038/nnano.2015.282} {\bibfield  {journal}
  {\bibinfo  {journal} {Nature Nanotechnology}\ ,\ \bibinfo {pages} {10}}
  (\bibinfo {year} {2015})}\BibitemShut {NoStop}%
\bibitem [{\citenamefont {Kubo}\ \emph {et~al.}(2011)\citenamefont {Kubo},
  \citenamefont {Grezes}, \citenamefont {Dewes}, \citenamefont {Umeda},
  \citenamefont {Isoya}, \citenamefont {Sumiya}, \citenamefont {Morishita},
  \citenamefont {Abe}, \citenamefont {Onoda}, \citenamefont {Ohshima},
  \citenamefont {Jacques}, \citenamefont {Dr{\'{e}}au}, \citenamefont {Roch},
  \citenamefont {Diniz}, \citenamefont {Auffeves}, \citenamefont {Vion},
  \citenamefont {Esteve},\ and\ \citenamefont {Bertet}}]{Kubo2011}%
  \BibitemOpen
  \bibfield  {author} {\bibinfo {author} {\bibfnamefont {Y.}~\bibnamefont
  {Kubo}}, \bibinfo {author} {\bibfnamefont {C.}~\bibnamefont {Grezes}},
  \bibinfo {author} {\bibfnamefont {A.}~\bibnamefont {Dewes}}, \bibinfo
  {author} {\bibfnamefont {T.}~\bibnamefont {Umeda}}, \bibinfo {author}
  {\bibfnamefont {J.}~\bibnamefont {Isoya}}, \bibinfo {author} {\bibfnamefont
  {H.}~\bibnamefont {Sumiya}}, \bibinfo {author} {\bibfnamefont
  {N.}~\bibnamefont {Morishita}}, \bibinfo {author} {\bibfnamefont
  {H.}~\bibnamefont {Abe}}, \bibinfo {author} {\bibfnamefont {S.}~\bibnamefont
  {Onoda}}, \bibinfo {author} {\bibfnamefont {T.}~\bibnamefont {Ohshima}},
  \bibinfo {author} {\bibfnamefont {V.}~\bibnamefont {Jacques}}, \bibinfo
  {author} {\bibfnamefont {A.}~\bibnamefont {Dr{\'{e}}au}}, \bibinfo {author}
  {\bibfnamefont {J.~F.}\ \bibnamefont {Roch}}, \bibinfo {author}
  {\bibfnamefont {I.}~\bibnamefont {Diniz}}, \bibinfo {author} {\bibfnamefont
  {A.}~\bibnamefont {Auffeves}}, \bibinfo {author} {\bibfnamefont
  {D.}~\bibnamefont {Vion}}, \bibinfo {author} {\bibfnamefont {D.}~\bibnamefont
  {Esteve}},\ and\ \bibinfo {author} {\bibfnamefont {P.}~\bibnamefont
  {Bertet}},\ }\bibfield  {title} {\bibinfo {title} {{Hybrid quantum circuit
  with a superconducting qubit coupled to a spin ensemble}},\ }\bibfield
  {journal} {\bibinfo  {journal} {Physical Review Letters}\ }\href
  {https://doi.org/10.1103/PhysRevLett.107.220501}
  {10.1103/PhysRevLett.107.220501} (\bibinfo {year} {2011})\BibitemShut
  {NoStop}%
\bibitem [{\citenamefont {Farr}\ \emph {et~al.}(2015)\citenamefont {Farr},
  \citenamefont {Goryachev}, \citenamefont {le~Floch}, \citenamefont {Bushev},\
  and\ \citenamefont {Tobar}}]{Farr2015}%
  \BibitemOpen
  \bibfield  {author} {\bibinfo {author} {\bibfnamefont {W.~G.}\ \bibnamefont
  {Farr}}, \bibinfo {author} {\bibfnamefont {M.}~\bibnamefont {Goryachev}},
  \bibinfo {author} {\bibfnamefont {J.-M.}\ \bibnamefont {le~Floch}}, \bibinfo
  {author} {\bibfnamefont {P.}~\bibnamefont {Bushev}},\ and\ \bibinfo {author}
  {\bibfnamefont {M.~E.}\ \bibnamefont {Tobar}},\ }\bibfield  {title} {\bibinfo
  {title} {{Evidence of dilute ferromagnetism in rare-earth doped yttrium
  aluminium garnet}},\ }\href {https://doi.org/10.1063/1.4931432} {\bibfield
  {journal} {\bibinfo  {journal} {Applied Physics Letters}\ }\textbf {\bibinfo
  {volume} {107}},\ \bibinfo {pages} {122401} (\bibinfo {year}
  {2015})}\BibitemShut {NoStop}%
\bibitem [{\citenamefont {Probst}\ \emph {et~al.}(2013)\citenamefont {Probst},
  \citenamefont {Rotzinger}, \citenamefont {W{\"{u}}nsch}, \citenamefont
  {Jung}, \citenamefont {Jerger}, \citenamefont {Siegel}, \citenamefont
  {Ustinov},\ and\ \citenamefont {Bushev}}]{Probst2013}%
  \BibitemOpen
  \bibfield  {author} {\bibinfo {author} {\bibfnamefont {S.}~\bibnamefont
  {Probst}}, \bibinfo {author} {\bibfnamefont {H.}~\bibnamefont {Rotzinger}},
  \bibinfo {author} {\bibfnamefont {S.}~\bibnamefont {W{\"{u}}nsch}}, \bibinfo
  {author} {\bibfnamefont {P.}~\bibnamefont {Jung}}, \bibinfo {author}
  {\bibfnamefont {M.}~\bibnamefont {Jerger}}, \bibinfo {author} {\bibfnamefont
  {M.}~\bibnamefont {Siegel}}, \bibinfo {author} {\bibfnamefont {A.~V.}\
  \bibnamefont {Ustinov}},\ and\ \bibinfo {author} {\bibfnamefont {P.~A.}\
  \bibnamefont {Bushev}},\ }\bibfield  {title} {\bibinfo {title} {{Anisotropic
  rare-earth spin ensemble strongly coupled to a superconducting resonator}},\
  }\href {https://doi.org/10.1103/PhysRevLett.110.157001} {\bibfield  {journal}
  {\bibinfo  {journal} {Physical Review Letters}\ }\textbf {\bibinfo {volume}
  {110}},\ \bibinfo {pages} {1} (\bibinfo {year} {2013})}\BibitemShut {NoStop}%
\bibitem [{\citenamefont {Zollitsch}\ \emph {et~al.}(2015)\citenamefont
  {Zollitsch}, \citenamefont {Mueller}, \citenamefont {Franke}, \citenamefont
  {Goennenwein}, \citenamefont {Brandt}, \citenamefont {Gross},\ and\
  \citenamefont {Huebl}}]{Zollitsch2015}%
  \BibitemOpen
  \bibfield  {author} {\bibinfo {author} {\bibfnamefont {C.~W.}\ \bibnamefont
  {Zollitsch}}, \bibinfo {author} {\bibfnamefont {K.}~\bibnamefont {Mueller}},
  \bibinfo {author} {\bibfnamefont {D.~P.}\ \bibnamefont {Franke}}, \bibinfo
  {author} {\bibfnamefont {S.~T.~B.}\ \bibnamefont {Goennenwein}}, \bibinfo
  {author} {\bibfnamefont {M.~S.}\ \bibnamefont {Brandt}}, \bibinfo {author}
  {\bibfnamefont {R.}~\bibnamefont {Gross}},\ and\ \bibinfo {author}
  {\bibfnamefont {H.}~\bibnamefont {Huebl}},\ }\bibfield  {title} {\bibinfo
  {title} {{High cooperativity coupling between a phosphorus donor spin
  ensemble and a superconducting microwave resonator}},\ }\href
  {https://doi.org/10.1063/1.4932658} {\bibfield  {journal} {\bibinfo
  {journal} {Applied Physics Letters}\ }\textbf {\bibinfo {volume} {107}},\
  \bibinfo {pages} {142105} (\bibinfo {year} {2015})}\BibitemShut {NoStop}%
\bibitem [{\citenamefont {Schuster}\ \emph {et~al.}(2010)\citenamefont
  {Schuster}, \citenamefont {Sears}, \citenamefont {Ginossar}, \citenamefont
  {Dicarlo}, \citenamefont {Frunzio}, \citenamefont {Morton}, \citenamefont
  {Wu}, \citenamefont {Briggs}, \citenamefont {Buckley}, \citenamefont
  {Awschalom},\ and\ \citenamefont {Schoelkopf}}]{Schuster2010}%
  \BibitemOpen
  \bibfield  {author} {\bibinfo {author} {\bibfnamefont {D.~I.}\ \bibnamefont
  {Schuster}}, \bibinfo {author} {\bibfnamefont {A.~P.}\ \bibnamefont {Sears}},
  \bibinfo {author} {\bibfnamefont {E.}~\bibnamefont {Ginossar}}, \bibinfo
  {author} {\bibfnamefont {L.}~\bibnamefont {Dicarlo}}, \bibinfo {author}
  {\bibfnamefont {L.}~\bibnamefont {Frunzio}}, \bibinfo {author} {\bibfnamefont
  {J.~J.~L.}\ \bibnamefont {Morton}}, \bibinfo {author} {\bibfnamefont
  {H.}~\bibnamefont {Wu}}, \bibinfo {author} {\bibfnamefont {G.~A.~D.}\
  \bibnamefont {Briggs}}, \bibinfo {author} {\bibfnamefont {B.~B.}\
  \bibnamefont {Buckley}}, \bibinfo {author} {\bibfnamefont {D.~D.}\
  \bibnamefont {Awschalom}},\ and\ \bibinfo {author} {\bibfnamefont {R.~J.}\
  \bibnamefont {Schoelkopf}},\ }\bibfield  {title} {\bibinfo {title}
  {{High-cooperativity coupling of electron-spin ensembles to superconducting
  cavities}},\ }\bibfield  {journal} {\bibinfo  {journal} {Physical Review
  Letters}\ }\href {https://doi.org/10.1103/PhysRevLett.105.140501}
  {10.1103/PhysRevLett.105.140501} (\bibinfo {year} {2010}),\ \Eprint
  {https://arxiv.org/abs/1006.0242} {arXiv:1006.0242} \BibitemShut {NoStop}%
\bibitem [{\citenamefont {Burnett}\ \emph {et~al.}(2014)\citenamefont
  {Burnett}, \citenamefont {Faoro}, \citenamefont {Wisby}, \citenamefont
  {Gurtovoi}, \citenamefont {Chernykh}, \citenamefont {Mikhailov},
  \citenamefont {Tulin}, \citenamefont {Shaikhaidarov}, \citenamefont
  {Antonov}, \citenamefont {Meeson}, \citenamefont {Tzalenchuk},\ and\
  \citenamefont {Lindstr{\"{o}}m}}]{Burnett2014}%
  \BibitemOpen
  \bibfield  {author} {\bibinfo {author} {\bibfnamefont {J.}~\bibnamefont
  {Burnett}}, \bibinfo {author} {\bibfnamefont {L.}~\bibnamefont {Faoro}},
  \bibinfo {author} {\bibfnamefont {I.}~\bibnamefont {Wisby}}, \bibinfo
  {author} {\bibfnamefont {V.~L.}\ \bibnamefont {Gurtovoi}}, \bibinfo {author}
  {\bibfnamefont {a.~V.}\ \bibnamefont {Chernykh}}, \bibinfo {author}
  {\bibfnamefont {G.~M.}\ \bibnamefont {Mikhailov}}, \bibinfo {author}
  {\bibfnamefont {V.~a.}\ \bibnamefont {Tulin}}, \bibinfo {author}
  {\bibfnamefont {R.}~\bibnamefont {Shaikhaidarov}}, \bibinfo {author}
  {\bibfnamefont {V.}~\bibnamefont {Antonov}}, \bibinfo {author} {\bibfnamefont
  {P.~J.}\ \bibnamefont {Meeson}}, \bibinfo {author} {\bibfnamefont {a.~Y.}\
  \bibnamefont {Tzalenchuk}},\ and\ \bibinfo {author} {\bibfnamefont
  {T.}~\bibnamefont {Lindstr{\"{o}}m}},\ }\bibfield  {title} {\bibinfo {title}
  {{Evidence for interacting two-level systems from the 1/f noise of a
  superconducting resonator.}},\ }\href {https://doi.org/10.1038/ncomms5119}
  {\bibfield  {journal} {\bibinfo  {journal} {Nature Communications}\ }\textbf
  {\bibinfo {volume} {5}},\ \bibinfo {pages} {4119} (\bibinfo {year}
  {2014})}\BibitemShut {NoStop}%
\bibitem [{\citenamefont {Oliver}\ and\ \citenamefont
  {Welander}(2013)}]{Oliver2013}%
  \BibitemOpen
  \bibfield  {author} {\bibinfo {author} {\bibfnamefont {W.~D.}\ \bibnamefont
  {Oliver}}\ and\ \bibinfo {author} {\bibfnamefont {P.~B.}\ \bibnamefont
  {Welander}},\ }\bibfield  {title} {\bibinfo {title} {{Materials in
  superconducting quantum bits}},\ }\href
  {https://doi.org/10.1557/mrs.2013.229} {\bibfield  {journal} {\bibinfo
  {journal} {MRS Bulletin}\ }\textbf {\bibinfo {volume} {38}},\ \bibinfo
  {pages} {816} (\bibinfo {year} {2013})}\BibitemShut {NoStop}%
\bibitem [{\citenamefont {Shoudu}\ \emph {et~al.}(1999)\citenamefont {Shoudu},
  \citenamefont {Siting}, \citenamefont {Xingda}, \citenamefont {Haobing},
  \citenamefont {Heyu}, \citenamefont {Shunxing},\ and\ \citenamefont
  {Jun}}]{Shoudu1999}%
  \BibitemOpen
  \bibfield  {author} {\bibinfo {author} {\bibfnamefont {Z.}~\bibnamefont
  {Shoudu}}, \bibinfo {author} {\bibfnamefont {W.}~\bibnamefont {Siting}},
  \bibinfo {author} {\bibfnamefont {S.}~\bibnamefont {Xingda}}, \bibinfo
  {author} {\bibfnamefont {W.}~\bibnamefont {Haobing}}, \bibinfo {author}
  {\bibfnamefont {Z.}~\bibnamefont {Heyu}}, \bibinfo {author} {\bibfnamefont
  {Z.}~\bibnamefont {Shunxing}},\ and\ \bibinfo {author} {\bibfnamefont
  {X.}~\bibnamefont {Jun}},\ }\bibfield  {title} {\bibinfo {title}
  {{Czochralski growth of rare-earth orthosilicates Y2SiO5 single crystals}},\
  }\href {https://doi.org/10.1016/S0022-0248(98)00553-3} {\bibfield  {journal}
  {\bibinfo  {journal} {Journal of Crystal Growth}\ }\textbf {\bibinfo {volume}
  {197}},\ \bibinfo {pages} {901} (\bibinfo {year} {1999})}\BibitemShut
  {NoStop}%
\bibitem [{\citenamefont {B{\"{a}}rnthaler}\ \emph {et~al.}(2010)\citenamefont
  {B{\"{a}}rnthaler}, \citenamefont {Rotter}, \citenamefont {Libisch},
  \citenamefont {Burgd{\"{o}}rfer}, \citenamefont {Gehler}, \citenamefont
  {Kuhl},\ and\ \citenamefont {St{\"{o}}ckmann}}]{Barnthaler2010}%
  \BibitemOpen
  \bibfield  {author} {\bibinfo {author} {\bibfnamefont {A.}~\bibnamefont
  {B{\"{a}}rnthaler}}, \bibinfo {author} {\bibfnamefont {S.}~\bibnamefont
  {Rotter}}, \bibinfo {author} {\bibfnamefont {F.}~\bibnamefont {Libisch}},
  \bibinfo {author} {\bibfnamefont {J.}~\bibnamefont {Burgd{\"{o}}rfer}},
  \bibinfo {author} {\bibfnamefont {S.}~\bibnamefont {Gehler}}, \bibinfo
  {author} {\bibfnamefont {U.}~\bibnamefont {Kuhl}},\ and\ \bibinfo {author}
  {\bibfnamefont {H.~J.}\ \bibnamefont {St{\"{o}}ckmann}},\ }\bibfield  {title}
  {\bibinfo {title} {{Probing decoherence through fano resonances}},\ }\href
  {https://doi.org/10.1103/PhysRevLett.105.056801} {\bibfield  {journal}
  {\bibinfo  {journal} {Physical Review Letters}\ }\textbf {\bibinfo {volume}
  {105}},\ \bibinfo {pages} {1} (\bibinfo {year} {2010})}\BibitemShut {NoStop}%
\bibitem [{\citenamefont {Pappas}\ \emph {et~al.}(2011)\citenamefont {Pappas},
  \citenamefont {Vissers}, \citenamefont {Wisbey}, \citenamefont {Kline},\ and\
  \citenamefont {Gao}}]{Pappas2011}%
  \BibitemOpen
  \bibfield  {author} {\bibinfo {author} {\bibfnamefont {D.~P.}\ \bibnamefont
  {Pappas}}, \bibinfo {author} {\bibfnamefont {M.~R.}\ \bibnamefont {Vissers}},
  \bibinfo {author} {\bibfnamefont {D.~S.}\ \bibnamefont {Wisbey}}, \bibinfo
  {author} {\bibfnamefont {J.~S.}\ \bibnamefont {Kline}},\ and\ \bibinfo
  {author} {\bibfnamefont {J.}~\bibnamefont {Gao}},\ }\bibfield  {title}
  {\bibinfo {title} {{Two level system loss in superconducting microwave
  resonators}},\ }\href {https://doi.org/10.1109/TASC.2010.2097578} {\bibfield
  {journal} {\bibinfo  {journal} {IEEE Transactions on Applied
  Superconductivity}\ }\textbf {\bibinfo {volume} {21}},\ \bibinfo {pages}
  {871} (\bibinfo {year} {2011})}\BibitemShut {NoStop}%
\bibitem [{\citenamefont {Faoro}\ and\ \citenamefont
  {Ioffe}(2006)}]{Faoro2006}%
  \BibitemOpen
  \bibfield  {author} {\bibinfo {author} {\bibfnamefont {L.}~\bibnamefont
  {Faoro}}\ and\ \bibinfo {author} {\bibfnamefont {L.~B.}\ \bibnamefont
  {Ioffe}},\ }\bibfield  {title} {\bibinfo {title} {{Quantum two level systems
  and kondo-like traps as possible sources of decoherence in superconducting
  qubits}},\ }\href {https://doi.org/10.1103/PhysRevLett.96.047001} {\bibfield
  {journal} {\bibinfo  {journal} {Physical Review Letters}\ }\textbf {\bibinfo
  {volume} {96}},\ \bibinfo {pages} {1} (\bibinfo {year} {2006})}\BibitemShut
  {NoStop}%
\bibitem [{\citenamefont {Maier-Flaig}(2013)}]{Maier-Flaig2013}%
  \BibitemOpen
  \bibfield  {author} {\bibinfo {author} {\bibfnamefont {H.}~\bibnamefont
  {Maier-Flaig}},\ }\emph {\bibinfo {title} {{Electron and nuclear spin
  properties of 145Neodymium doped Y2SiO5}}},\ \href@noop {} {Master's
  thesis},\ \bibinfo  {school} {Karlsrhue Institute of Technology} (\bibinfo
  {year} {2013})\BibitemShut {NoStop}%
\bibitem [{\citenamefont {Wisby}\ \emph {et~al.}(2016)\citenamefont {Wisby},
  \citenamefont {de~Graaf}, \citenamefont {Gwilliam}, \citenamefont {Adamyan},
  \citenamefont {Kubatkin}, \citenamefont {Meeson}, \citenamefont
  {Tzalenchuk},\ and\ \citenamefont {Lindstr{\"{o}}m}}]{Wisby2016}%
  \BibitemOpen
  \bibfield  {author} {\bibinfo {author} {\bibfnamefont {I.~S.}\ \bibnamefont
  {Wisby}}, \bibinfo {author} {\bibfnamefont {S.~E.}\ \bibnamefont {de~Graaf}},
  \bibinfo {author} {\bibfnamefont {R.}~\bibnamefont {Gwilliam}}, \bibinfo
  {author} {\bibfnamefont {A.}~\bibnamefont {Adamyan}}, \bibinfo {author}
  {\bibfnamefont {S.~E.}\ \bibnamefont {Kubatkin}}, \bibinfo {author}
  {\bibfnamefont {P.~J.}\ \bibnamefont {Meeson}}, \bibinfo {author}
  {\bibfnamefont {A.~Y.}\ \bibnamefont {Tzalenchuk}},\ and\ \bibinfo {author}
  {\bibfnamefont {T.}~\bibnamefont {Lindstr{\"{o}}m}},\ }\bibfield  {title}
  {\bibinfo {title} {{Angle-Dependent Microresonator ESR Characterization of
  Locally Doped Gd3+:Al2O3}},\ }\href
  {https://doi.org/10.1103/PhysRevApplied.6.024021} {\bibfield  {journal}
  {\bibinfo  {journal} {Physical Review Applied}\ }\textbf {\bibinfo {volume}
  {6}},\ \bibinfo {pages} {024021} (\bibinfo {year} {2016})}\BibitemShut
  {NoStop}%
\bibitem [{\citenamefont {Abe}\ \emph {et~al.}(2011)\citenamefont {Abe},
  \citenamefont {Wu}, \citenamefont {Ardavan},\ and\ \citenamefont
  {Morton}}]{Abe2011}%
  \BibitemOpen
  \bibfield  {author} {\bibinfo {author} {\bibfnamefont {E.}~\bibnamefont
  {Abe}}, \bibinfo {author} {\bibfnamefont {H.}~\bibnamefont {Wu}}, \bibinfo
  {author} {\bibfnamefont {A.}~\bibnamefont {Ardavan}},\ and\ \bibinfo {author}
  {\bibfnamefont {J.~J.~L.}\ \bibnamefont {Morton}},\ }\bibfield  {title}
  {\bibinfo {title} {{Electron spin ensemble strongly coupled to a
  three-dimensional microwave cavity}},\ }\href
  {https://doi.org/10.1063/1.3601930} {\bibfield  {journal} {\bibinfo
  {journal} {Applied Physics Letters}\ }\textbf {\bibinfo {volume} {98}},\
  \bibinfo {pages} {3} (\bibinfo {year} {2011})}\BibitemShut {NoStop}%
\bibitem [{\citenamefont {Bienfait}\ \emph {et~al.}(2016)\citenamefont
  {Bienfait}, \citenamefont {Pla}, \citenamefont {Kubo}, \citenamefont {Zhou},
  \citenamefont {Stern}, \citenamefont {Lo}, \citenamefont {Weis},
  \citenamefont {Schenkel}, \citenamefont {Vion}, \citenamefont {Esteve},
  \citenamefont {Morton},\ and\ \citenamefont {Bertet}}]{Bienfait2015a}%
  \BibitemOpen
  \bibfield  {author} {\bibinfo {author} {\bibfnamefont {A.}~\bibnamefont
  {Bienfait}}, \bibinfo {author} {\bibfnamefont {J.~J.}\ \bibnamefont {Pla}},
  \bibinfo {author} {\bibfnamefont {Y.}~\bibnamefont {Kubo}}, \bibinfo {author}
  {\bibfnamefont {X.}~\bibnamefont {Zhou}}, \bibinfo {author} {\bibfnamefont
  {M.}~\bibnamefont {Stern}}, \bibinfo {author} {\bibfnamefont {C.~C.}\
  \bibnamefont {Lo}}, \bibinfo {author} {\bibfnamefont {C.~D.}\ \bibnamefont
  {Weis}}, \bibinfo {author} {\bibfnamefont {T.}~\bibnamefont {Schenkel}},
  \bibinfo {author} {\bibfnamefont {D.}~\bibnamefont {Vion}}, \bibinfo {author}
  {\bibfnamefont {D.}~\bibnamefont {Esteve}}, \bibinfo {author} {\bibfnamefont
  {J.~J.~L.}\ \bibnamefont {Morton}},\ and\ \bibinfo {author} {\bibfnamefont
  {P.}~\bibnamefont {Bertet}},\ }\bibfield  {title} {\bibinfo {title}
  {{Controlling spin relaxation with a cavity}},\ }\href
  {https://doi.org/10.1038/nature16944} {\bibfield  {journal} {\bibinfo
  {journal} {Nature}\ }\textbf {\bibinfo {volume} {531}},\ \bibinfo {pages}
  {74} (\bibinfo {year} {2016})}\BibitemShut {NoStop}%
\bibitem [{\citenamefont {Ortu}\ \emph {et~al.}(2018)\citenamefont {Ortu},
  \citenamefont {Tiranov}, \citenamefont {Welinski}, \citenamefont
  {Fr{\"{o}}wis}, \citenamefont {Gisin}, \citenamefont {Ferrier}, \citenamefont
  {Goldner},\ and\ \citenamefont {Afzelius}}]{Ortu2018}%
  \BibitemOpen
  \bibfield  {author} {\bibinfo {author} {\bibfnamefont {A.}~\bibnamefont
  {Ortu}}, \bibinfo {author} {\bibfnamefont {A.}~\bibnamefont {Tiranov}},
  \bibinfo {author} {\bibfnamefont {S.}~\bibnamefont {Welinski}}, \bibinfo
  {author} {\bibfnamefont {F.}~\bibnamefont {Fr{\"{o}}wis}}, \bibinfo {author}
  {\bibfnamefont {N.}~\bibnamefont {Gisin}}, \bibinfo {author} {\bibfnamefont
  {A.}~\bibnamefont {Ferrier}}, \bibinfo {author} {\bibfnamefont
  {P.}~\bibnamefont {Goldner}},\ and\ \bibinfo {author} {\bibfnamefont
  {M.}~\bibnamefont {Afzelius}},\ }\bibfield  {title} {\bibinfo {title}
  {{Simultaneous coherence enhancement of optical and microwave transitions in
  solid-state electronic spins}},\ }\href
  {https://doi.org/10.1038/s41563-018-0138-x} {\bibfield  {journal} {\bibinfo
  {journal} {Nature Materials}\ }\textbf {\bibinfo {volume} {17}},\ \bibinfo
  {pages} {671} (\bibinfo {year} {2018})}\BibitemShut {NoStop}%
\bibitem [{\citenamefont {Kukharchyk}\ \emph {et~al.}(2014)\citenamefont
  {Kukharchyk}, \citenamefont {Pal}, \citenamefont {R{\"{o}}diger},
  \citenamefont {Ludwig}, \citenamefont {Probst}, \citenamefont {Ustinov},
  \citenamefont {Bushev},\ and\ \citenamefont {Wieck}}]{Kukharchyk2014}%
  \BibitemOpen
  \bibfield  {author} {\bibinfo {author} {\bibfnamefont {N.}~\bibnamefont
  {Kukharchyk}}, \bibinfo {author} {\bibfnamefont {S.}~\bibnamefont {Pal}},
  \bibinfo {author} {\bibfnamefont {J.}~\bibnamefont {R{\"{o}}diger}}, \bibinfo
  {author} {\bibfnamefont {A.}~\bibnamefont {Ludwig}}, \bibinfo {author}
  {\bibfnamefont {S.}~\bibnamefont {Probst}}, \bibinfo {author} {\bibfnamefont
  {A.~V.}\ \bibnamefont {Ustinov}}, \bibinfo {author} {\bibfnamefont
  {P.}~\bibnamefont {Bushev}},\ and\ \bibinfo {author} {\bibfnamefont {A.~D.}\
  \bibnamefont {Wieck}},\ }\bibfield  {title} {\bibinfo {title}
  {{Photoluminescence of focused ion beam implanted Er3+:Y2SiO5 crystals}},\
  }\href {https://doi.org/10.1002/pssr.201409304} {\bibfield  {journal}
  {\bibinfo  {journal} {physica status solidi (RRL) - Rapid Research Letters}\
  }\textbf {\bibinfo {volume} {8}},\ \bibinfo {pages} {880} (\bibinfo {year}
  {2014})}\BibitemShut {NoStop}%
\bibitem [{\citenamefont {Welinski}\ \emph {et~al.}(2016)\citenamefont
  {Welinski}, \citenamefont {Ferrier}, \citenamefont {Afzelius},\ and\
  \citenamefont {Goldner}}]{Welinski2016}%
  \BibitemOpen
  \bibfield  {author} {\bibinfo {author} {\bibfnamefont {S.}~\bibnamefont
  {Welinski}}, \bibinfo {author} {\bibfnamefont {A.}~\bibnamefont {Ferrier}},
  \bibinfo {author} {\bibfnamefont {M.}~\bibnamefont {Afzelius}},\ and\
  \bibinfo {author} {\bibfnamefont {P.}~\bibnamefont {Goldner}},\ }\bibfield
  {title} {\bibinfo {title} {{High-resolution optical spectroscopy and magnetic
  properties of Yb3+ in Y2SiO5}},\ }\href
  {https://doi.org/10.1103/PhysRevB.94.155116} {\bibfield  {journal} {\bibinfo
  {journal} {Physical Review B}\ }\textbf {\bibinfo {volume} {94}},\ \bibinfo
  {pages} {155116} (\bibinfo {year} {2016})}\BibitemShut {NoStop}%
\end{thebibliography}%

\end{document}